\documentclass[preprint,12pt]{elsarticle}

\usepackage{amsmath}
\usepackage{amssymb} 
\usepackage{bm}

\newcommand{\x}[1]{{\bf x}_{#1}}

\journal{Optics Communications}

\begin{document}

\begin{frontmatter}



\title{Ptychography by changing the area of probe light and scaled ptychography}


\author[a]{Tomoyoshi Shimobaba\corref{cor1}}
\cortext[cor1]{Tel: +81 43 290 3361; fax: +81 43 290 3361}
\ead{shimobaba@faculty.chiba-u.jp}
\author[a]{Takashi Kakue}
\author[a]{Naohisa Okada}
\author[a]{Yutaka Endo}
\author[a]{Ryuji Hirayama}
\author[a]{Daisuke Hiyama}
\author[a]{Tomoyoshi Ito}

\address[a]{Chiba University, Graduate School of Engineering, 1--33 Yayoi--cho, Inage--ku, Chiba, Japan, 263--8522}

\begin{abstract}
Ptychography is a promising phase retrieval technique for visible light, X-ray and electron beams.
Conventional ptychography reconstructs the amplitude and phase of an object light from a set of the diffraction intensity patterns obtained by the X-Y moving of the probe light.
The X-Y moving of the probe light requires two control parameters and accuracy of the locations.
We propose ptychography by changing the area of the probe light using  only one control parameter,  instead of the X-Y moving of the probe light.
The proposed method has faster convergence speed.
In addition, we propose scaled ptychography using scaled diffraction calculation in order to magnify retrieved object lights clearly.
\end{abstract}

\begin{keyword}
Phase retrieval \sep Ptychography \sep Scaled diffraction

\end{keyword}

\end{frontmatter}

\section{Introduction}
Various coherent diffractive imaging (CDI) methods have been proposed. 
Ptychography is a CDI method that was invented by Hoppe\cite{hoppe1969}.
Ptychography is a promising phase retrieval technique for visible light \cite{ maiden2010}, X-ray \cite{thibault2008} and electron beams \cite{hue2011}.
Conventional ptychography reconstructs the amplitude and phase of an object light from a set of diffraction intensity patterns at different lateral positions using the X-Y moving of the probe light.

As reconstruction algorithm, the ptychographical iterative engine (PIE) \cite{rodenburg2004} has been widely used.
The probe light in PIE must be accurately determined in advance. 
Furthermore, the extended PIE (ePIE) algorithm, which can guess the complex amplitudes of an object light and probe simultaneously \cite{maiden2009}.
Instead of the X-Y moving of the probe light using two control parameters,   a rotating probe light with a diffuser has been proposed recently \cite{wang2014}.
By reducing the movement to only one control parameter (rotating angle) unlike the X-Y moving, the control error of the X-Y moving that degrades the reconstruction quality can be suppressed \cite{wang2014}.

In this paper, we propose ptychography by changing the area of the probe light instead of the X-Y moving of the probe light.
The proposed method also has only one control parameter (the area of probe light).
In addition, scaled ptychography using scaled diffraction calculation, e.g. Refs. \cite{scale1,scale2,scale3, scale4, scale5, scale6, scale7, scale8, scale9}, is presented in order to magnify a retrieved object clearly.
We show the effectiveness through computer simulation.

\section{Ptychography by changing the area of probe light}
\begin{figure}[htb]
\centerline{
\includegraphics[width=6cm]{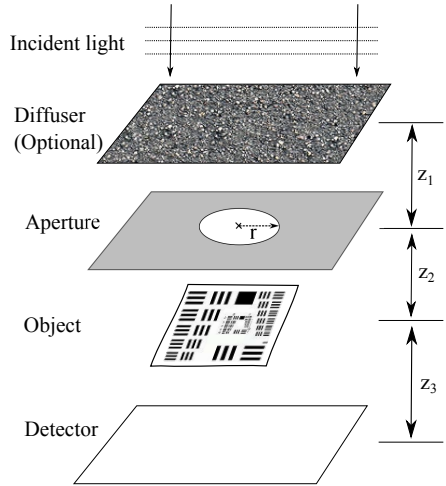}}
\caption{Setup of the optical system. The diffuser is optional.}
\label{fig:system}
\end{figure}

We assume the optical system as shown in Fig.\ref{fig:system}.
A planar or spherical light can be used as the incident light with a wavelength of $\lambda$.
Although a diffuser in the rotating probe light \cite{wang2014} is necessary, a diffuser in the proposed method is optional. 
The incident light illuminates the aperture directly or via through the diffuser.
In order to change the area of the probe light, the aperture can control the radius $r$ of the hole.
The probe light illuminates an object and the exit light travels to the detector.
We can obtain a set of diffraction intensity patterns by changing the area of the probe light.

The procedure of the ptychographic reconstruction by changing the area of the probe light is almost the same as the original PIE  \cite{rodenburg2004} except for changing the area of the probe light.
The procedure is as follows:

(1) We randomly select one radius $r$ of the aperture from the predefined 16 radii, and record the diffraction intensity pattern $I({\bm x_3})$ using a detector such as a CCD camera.
${\bm x_3}$ is the position vector on the detector plane.
We define the 16 radii as $r \in [128 \mu m, 144 \mu m \cdots 3,840 \mu m]$ at the intervals of $256 \mu$m.

(2) The exit light $\psi_j({\bm x_2})$, where the subscript $j$ denotes $j$-th iteration, is expressed by 
\begin{equation}
\psi_j({\bm x_2})=O_j({\bm x_2}) P({\bm x_2}),
\end{equation}
where $O_j({\bm x_2})$ is the guessed object light and $P({\bm x_2})$ is the probe light on the object plane.
${\bm x_2}$ is the position vectors on the object plane.
The initial $O_j({\bm x_2})$ is set to $O_j({\bm x_2})=1$.

(3) The new exit light $\psi'_j({\bm x_2})$ on the object plane is calculated by,
\begin{equation}
\psi'_j({\bm x_2})={\rm Prop}_{-z_3} [\sqrt{I({\bm x_3})} \exp(i \theta({\bm x_3})) ],
\end{equation}
where the operator $\rm Prop_z[\cdot]$ denotes the light propagation calculation with the propagation distance $z$, such as the angular spectrum method (ASM), Fresnel diffraction and Fraunhofer diffraction \cite{goodman}.
$\theta({\bm x_3})$ is the phase distribution of the propagated exit light $\phi({\bm x_3})={\rm Prop}_{z_3}[\psi_j({\bm x_2})]$ from $\psi_j({\bm x_2})$, that is, 
\begin{equation}
\theta({\bm x_3})=\tan^{-1} \frac{\Im[\phi({\bm x_3})]}{\Re[\phi({\bm x_3})]},
\end{equation}
where $\Re[\cdot]$ and $\Im[\cdot]$ take the real and imaginary values of the complex amplitude.

(4) The next guessed object light $O_{j+1}(\bm x_1)$ is calculated by the following update function of PIE:
\begin{equation}
O_{j+1}({\bm x_2})=
O_j({\bm x_2})+\frac{P^{*}({\bm x_2})}{|P({\bm x_2})|^2_{max}}
[\psi'({\bm x_2}) - \psi({\bm x_2})],
\end{equation}
where $*$ denotes the complex conjugate and $| \cdot |^2_{max}$ means the maximum value of $| \cdot |^2$.
We repeat the above steps (1) to (4) until a certain number of iterations is reached.

\section{Results}
\begin{figure}[htb]
\centerline{
\includegraphics[width=10cm]{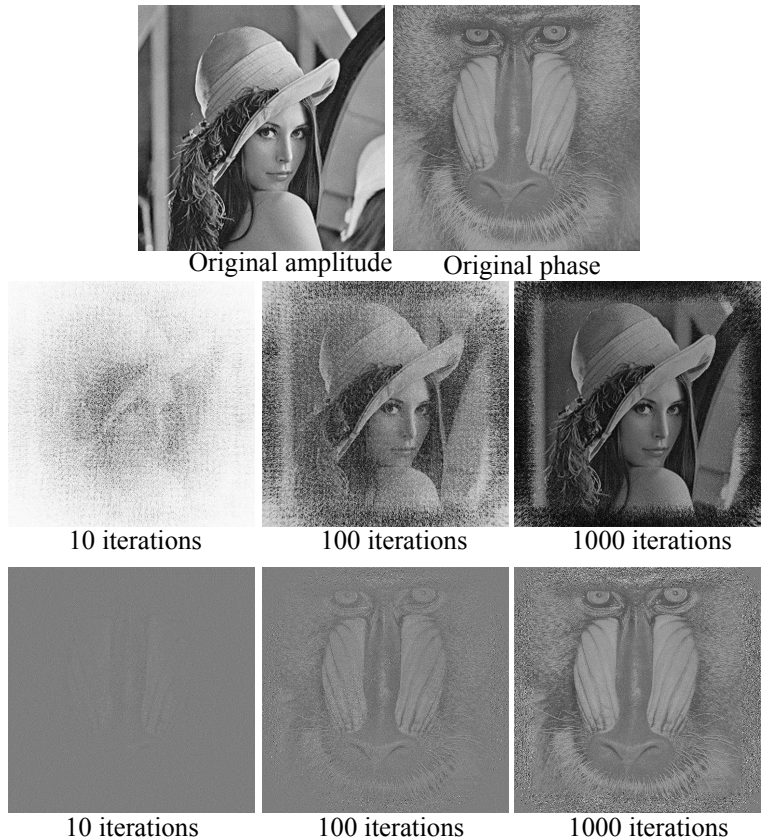}}
\caption{Ptychographic reconstruction by changing the area of the probe with the diffuser. The calculation conditions are $z_1=20$mm, $z_2=20$mm and $z_3=10$mm.}
\label{fig:reconst_rnd}
\end{figure}

\begin{figure}[htb]
\centerline{
\includegraphics[width=10cm]{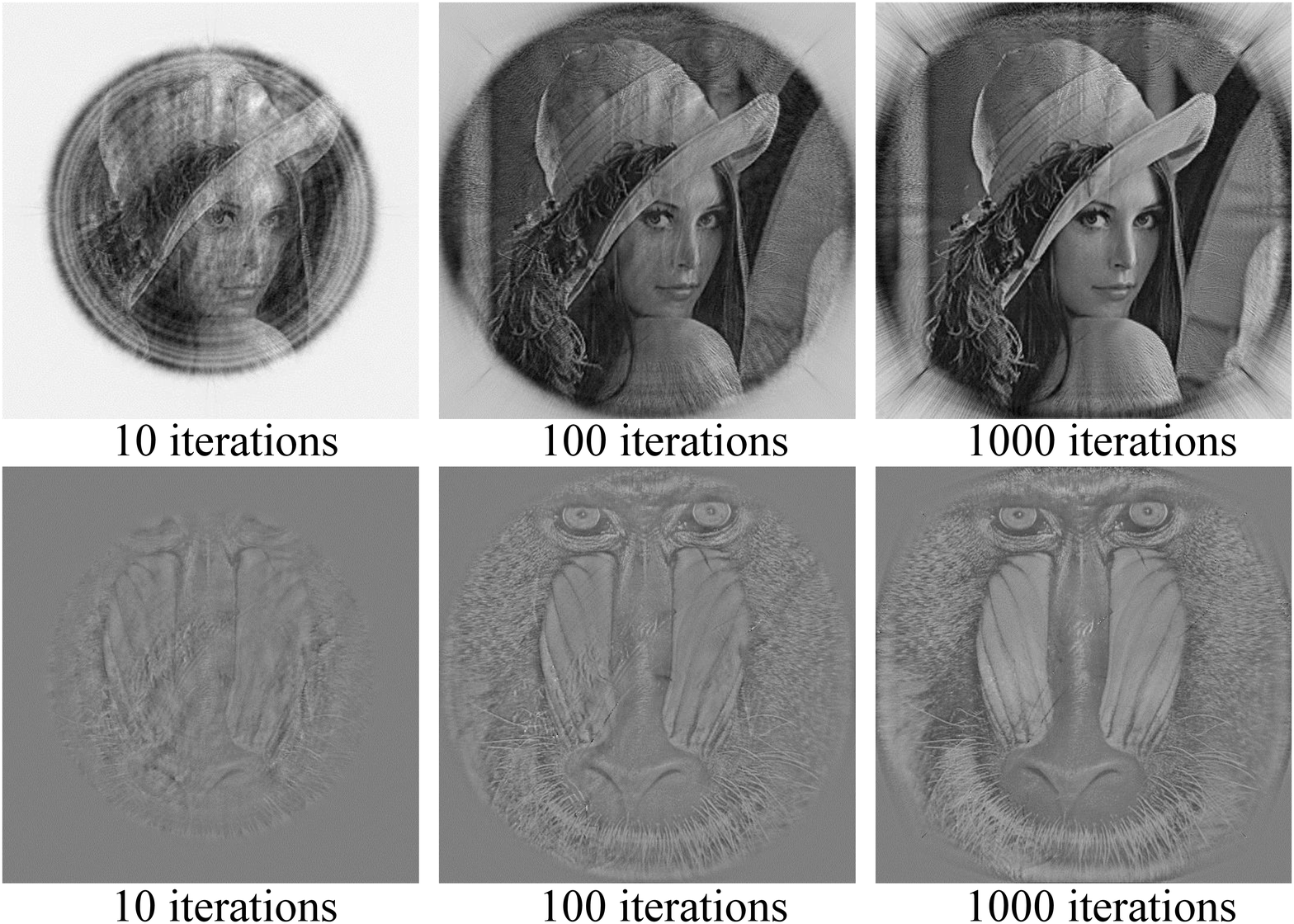}}
\caption{Ptychographic reconstruction by changing the area of the probe without the diffuser. The calculation conditions are $z_1=20$mm, $z_2=20$mm and $z_3=40$mm.}
\label{fig:reconst_without _rnd}
\end{figure}

We verify the procedure through simulation.
In Fig.\ref{fig:reconst_rnd}, we use ``Lena'' and ``Mandrill'' as the amplitude and phase of the original light.
The phase is linearly mapped at 0 and 255 pixel values of ``Mandrill'' to $-\pi$ and $+\pi$ radians, respectively.
Figures \ref{fig:reconst_rnd} and \ref{fig:reconst_without _rnd} show the retrieved amplitude and phase distributions by the proposed method with and without the diffuser, respectively.
The number of the iterations is 10, 100 and 1,000, respectively.
In the simulation, we generate the probe light on the object plane as follows:
\begin{equation}
P({\bm x_2})={\rm Prop}_{z_2}[ {\rm Circ}(\frac{|{\bm x_1}|}{r_j}) \cdot {\rm Prop}_{z_1}[{\rm Rnd}({\bm x_0}) ] ],
\end{equation}
where ${\rm Rnd}({\bm x_0})$ denotes the generation of a random phase on the diffuser, and ${\rm Circ}(|{\bm x_1}|/r_j)$ denotes the aperture being capable of changing the area and ${\rm Circ}(|{\bm x_1}|/r_j)=1$ when $|{\bm x_1}|/r \le 1$, otherwise 0.

\begin{figure}[htb]
\centerline{
\includegraphics[width=10cm]{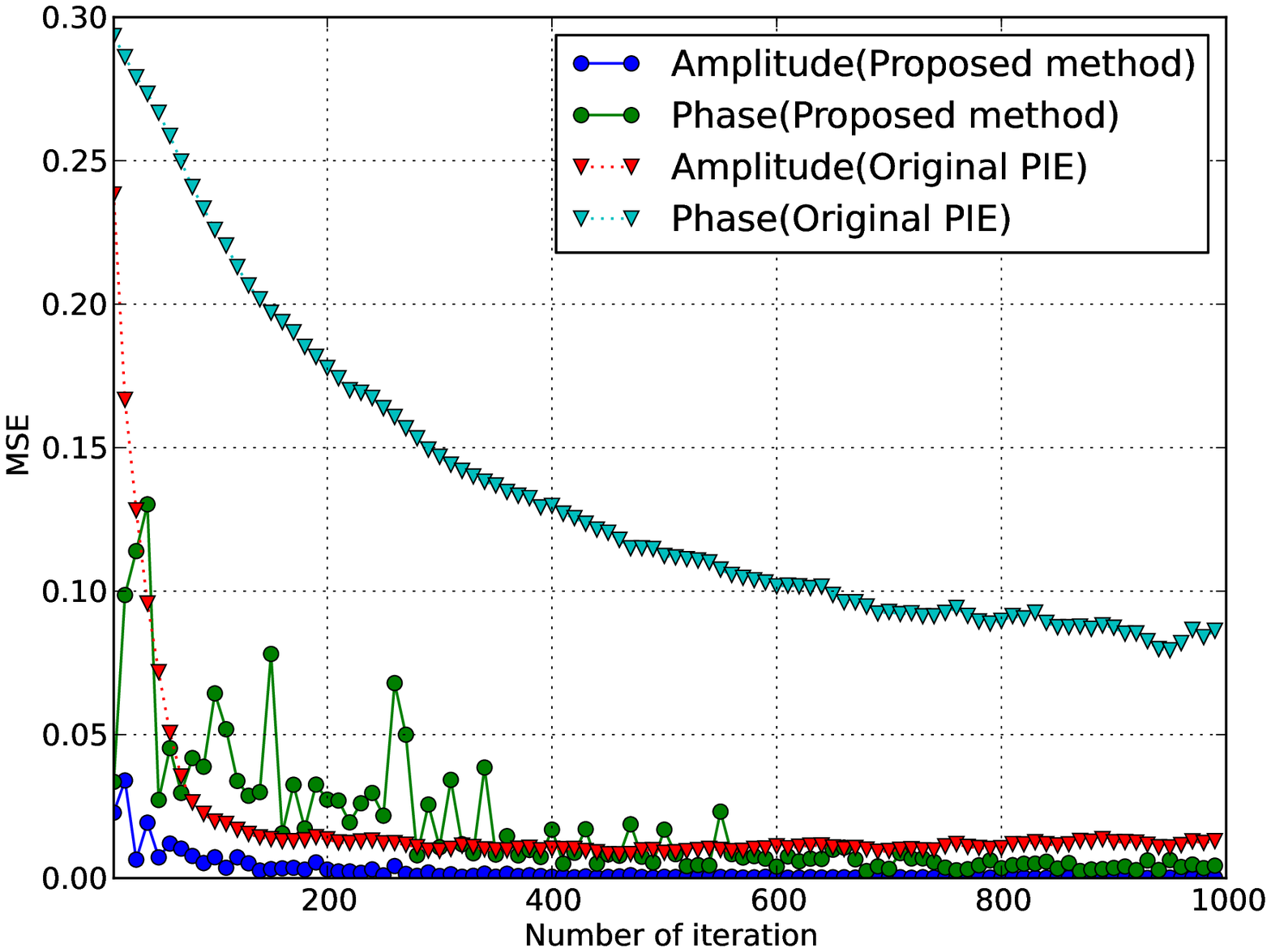}}
\caption{MSEs between the original and retrieved objects using the original PIE and proposed method. }
\label{fig:mse}
\end{figure}

We estimate the error of Fig.\ref{fig:reconst_rnd} between the amplitudes and phases of the original object $O(\bm x)$ and the retrieved object $O_j(\bm x)$ by the mean square error (MSE) that is calculated by ${\rm MSE}=\frac{1}{NM} \sum_{\bm x} |f(\bm x) - f_j(\bm x)|^2$ where $N$ and $M$ are the horizontal and vertical pixel numbers, and $f(\bm x)$ and $f_j(\bm x)$ are the amplitudes or phases of the original and retrieved objects, respectively.
The MSE indicates a better result when the value is low.
In the simulation, $M$ and $N$ are 512 pixels, and we calculate the MSE in the region of interest whose size is $400 \times 400$ pixels.
The calculation conditions are $\lambda=633$ nm, $z_1=20$ mm, $z_2=20$ mm, $z_3=10$ mm, and the sampling pitch on the object and detector is 8$\mu$m.
We use ASM as the light propagation.
For comparison, we show the MSE when using the original PIE \cite{rodenburg2004} that randomly moves the probe light in the lateral direction, and the calculation conditions are the same as those shown in Fig.\ref{fig:reconst_rnd}.
Overall, the proposed method quickly converges in both the amplitude and phase, as compared with the original PIE.
In these calculation conditions, the proposed method converges in  iterations of 800 and over.

Next, we discuss the subtopic, scaled ptychography.
Scaled ptychography observes a magnified retrieved object clearly.
If we want to observe a magnified retrieved object, in general, we use a digital magnification technique, such as linear and bicubic interpolations.
Instead of such digital magnification technique, we use scaled diffraction.
Scaled diffraction calculations have been proposed by many authors \cite{scale1,scale2,scale3, scale4, scale5, scale6, scale7, scale8, scale9}.
Here, we use scaled ASM \cite{scale8, nu} that can calculate ASM in different sampling pitches on the source and destination planes.

In order to verify the scaled ptychography, we prepare a reduced original object $O({\bm x_2})$.
We assume that the sampling pitch $p_o$ on the reduced original object is 2$\mu$m and the sampling pitch $p_d$ on the detector  is 8 $\mu$m.

\begin{figure}[htb]
\centerline{
\includegraphics[width=10cm]{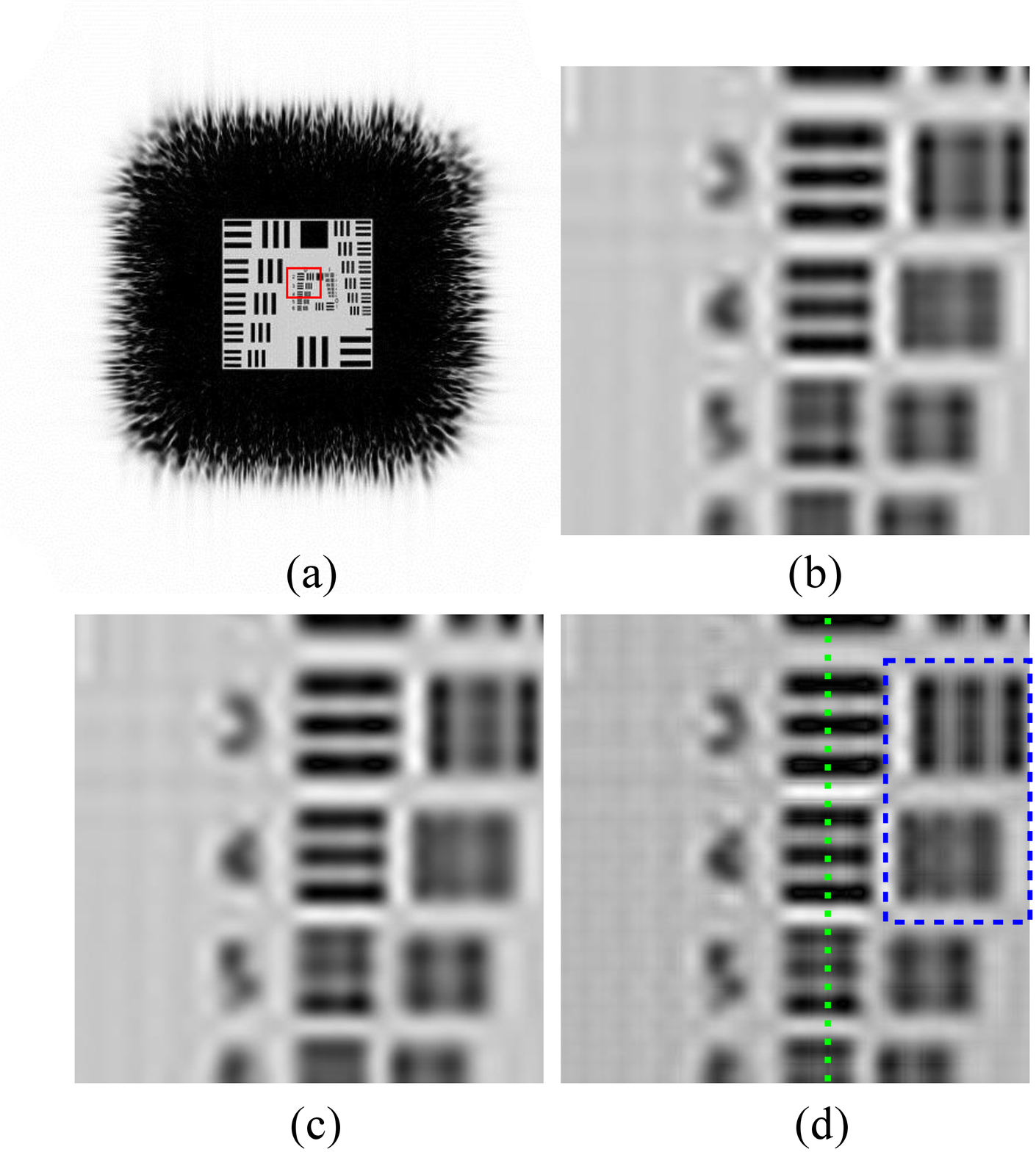}}
\caption{Ptychographic reconstructions (a) Normal reconstruction without magnification (b) Bicubic interpolation (c) Lanczos-3 interpolation (d) Scaled ASM.}
\label{fig:scale}
\end{figure}

\begin{figure}[htb]
\centerline{
\includegraphics[width=10cm]{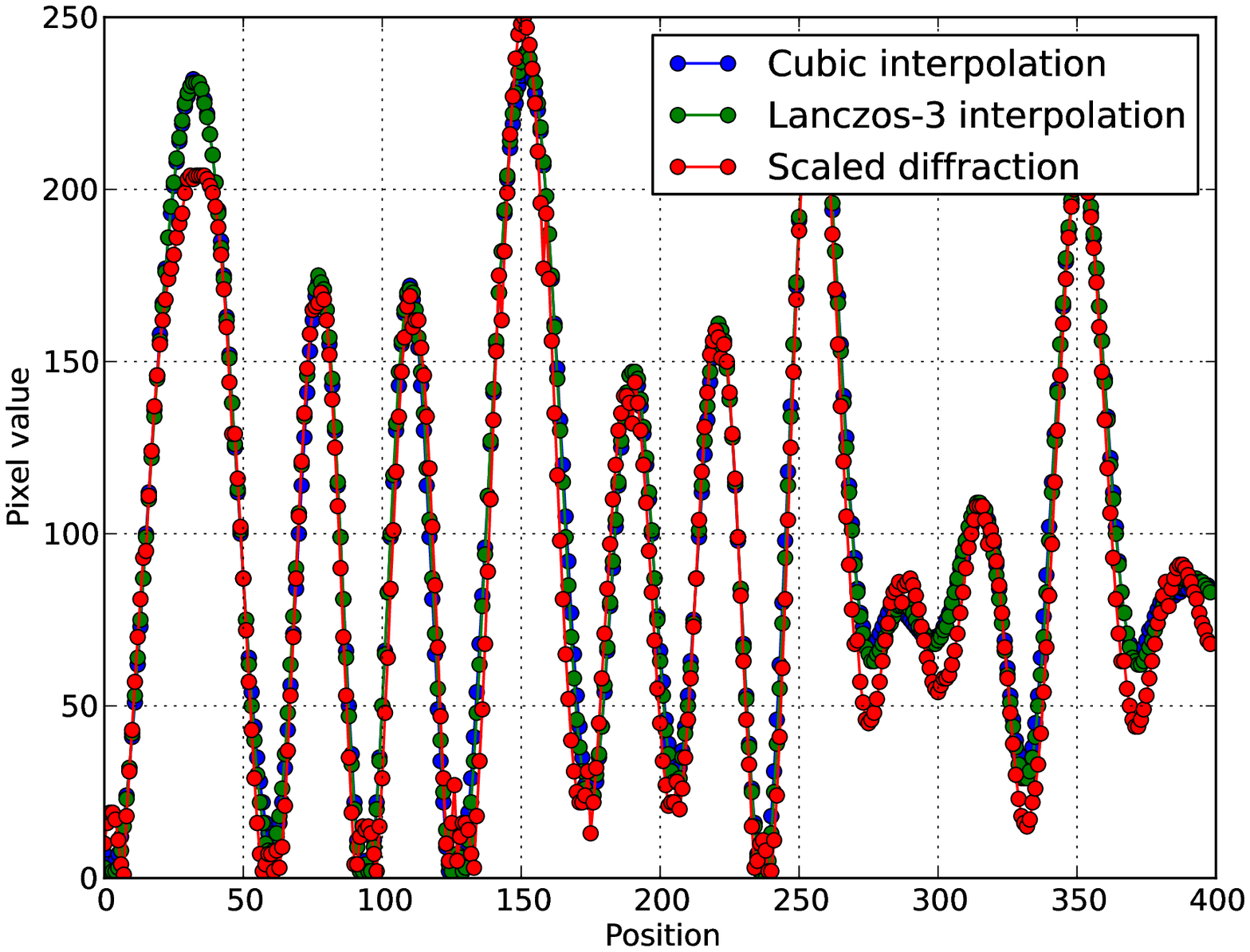}}
\caption{Pixel variation on the green dashed line of Fig.\ref{fig:scale}(d) and the corresponding parts in Figs.\ref{fig:scale}(b) and (c).}
\label{fig:line}
\end{figure}

Figure \ref{fig:scale}(a) shows the retrieved object $O_j({\bm x_2})$ without magnification by the proposed method with the diffuser.
The result shows that the retrieved object is reduced because $p_o$ is one-quarter that of $p_d$.
Figures \ref{fig:scale}(b) and (c) show the magnified objects in the red square of Fig. \ref{fig:scale}(a) by the bicubic and Lanczos-3 interpolations with 16x magnification.
Figure \ref{fig:scale}(d) shows the magnified images by the scaled ASM with 16x magnification.

The magnified object $O_j'({\bm x_2})$ from the retrieved object $O_j({\bm x_2})$ using the scaled ASM is obtained by,
\begin{equation}
O_j'({\bm x_2})={\rm Prop}_{-z'}^{p_d, p_d/16} [ {\rm Prop}_{z'} [O_j({\bm x_2}) ] ],
\end{equation}
where $z'$ is a short distance, and the scaled ASM operator is ${\rm Prop}_{z}^{ p_1, p_2}[\cdot]$ where the parameters $p_1$ and $p_2$ are the sampling rates on the source and destination planes, respectively.
The first propagation is normal ASM with the sampling pitch $p_d$.

The blue dashed square in Fig.\ref{fig:scale}(d) is well resolved, compared with the corresponding areas in Figs.\ref{fig:scale}(b) and (c).
Figure \ref{fig:line} shows the pixel variation on the green dashed line of Fig. \ref{fig:scale}(d) and the corresponding parts in Figs.\ref{fig:scale} (b) and (c).
The position from about 260 to 340 in Fig.\ref{fig:line} corresponds to  group 5 of the test target.
The scaled ASM is well resolved in the position from 260 to 340 because the pixel variation of the scaled ASM is larger than that of the others.   
We use our computational wave optics library, CWO++ \cite{cwo} in all the calculations above.

\section{Conclusion}
We conclude this work.
We propose ptychography by changing the area of the probe light.
The convergence speed of the proposed method is faster than the original PIE.
In addition, use of the diffuser in the proposed method is optional unlike the rotating diffuser method \cite{wang2014}.
We also propose scaled ptychography using scaled diffraction calculations.
The scaled method is well resolved, compared with digital magnification methods. 
We aim to apply the proposed method in a practical optical system. 

\section*{Acknowlegement}
This work is partially supported by JSPS KAKENHI Grant Numbers 25330125 and 25240015, and the Kayamori Foundation of Information Science Advancement and Yazaki Memorial Foundation for Science and Technology.

\bibliographystyle{model1a-num-names}
\bibliography{<your-bib-database>}

\begin{thebibliography}{99}


\bibitem{hoppe1969}
W. Hoppe,  ``Diffraction in inhomogeneous primary light fields. 1. Principle of phase determination from electron diffraction interference,''  Acta Crystallogr. A {\bf 25}, 495-–501 (1969).

\bibitem{maiden2010}
A. M. Maiden, J. M. Rodenburg, and M. J. Humphry, ``Optical ptychography: a practical implementation with useful resolution,'' Opt. Lett. {\bf 35}, 2585--2587 (2010).

\bibitem{thibault2008}
P. Thibault, M. Dierolf, A. Menzel, O. Bunk, C. David, and F. Pfeiffer,  ``High-Resolution Scanning X-ray Diffraction Microscopy,''  Science {\bf 321} (5887), 379--382 (2008).

\bibitem{hue2011}
F. H\"{u}e, J. M. Rodenburg, A. M. Maiden, and P. A. Midgley, ``Extended ptychography in the transmission electron microscope: Possibilities and limitations,'' Ultramicroscopy {\bf 111}, 1117--1123 (2011).

\bibitem{rodenburg2004}
J.M. Rodenburg and H.M.L. Faulkner,  ``A phase retrieval algorithm for shifting illumination,''  Appl. Phys. Lett. {\bf 85}, 4795–-4797 (2004).

\bibitem{maiden2009}
A. M. Maiden, J. M. Rodenburg  ``An improved ptychographical phase retrieval algorithm for diffractive imaging,'' Ultramicroscopy {\bf 109}, 1256--1262 (2009). 

\bibitem{wang2014}
H. Wang, C. Liu, X. Pan, J. Cheng, and J. Zhu,  ``Phase imaging with rotating illumination,''   Chin. Opt. Lett. {\bf 12}, 010501- (2014) .



\bibitem{scale1}
P. Ferraro, S. D. Nicola, G. Coppola, A. Finizio, D. Alfieri, and G. Pierattini, ``Controlling image size as a function of distance and wavelength in Fresnel-transform reconstruction of digital holograms,''  Opt. Lett. {\bf 29}, 854--856 (2004). 

\bibitem{scale2}
 L. Yaroslavsky, ``Optical transforms in digital holography,'' Proc. SPIE Holography 2005: International Conference on Holography, Optical Recording, and Processing of Information, {\bf 6252},  625216 (2006).

\bibitem{scale3}
R. P. Muffoletto, J. M. Tyler, and J. E. Tohline, ``Shifted Fresnel diffraction for computational holography,''  Opt. Express {\bf 15}, 5631--5640 (2007).


\bibitem{scale4}
M. Paturzo, P. Memmolo, A. Finizio, R. N{\"a}s{\"a}nen, T. J. Naughton, and P. Ferraro, ``Synthesis and display of dynamic holographic 3D scenes with real-world objects,'' Opt. Express {\bf 18}, 8806--8815 (2010).


\bibitem{scale5}
J. F. Restrepo and J. G. -Sucerquia, ``Magnified reconstruction of digitally recorded holograms by Fresnel-Bluestein transform,'' Appl. Opt. {\bf 49}, 6430--6435 (2010) .


\bibitem{scale6}
L. Bilevich and L. Yaroslavsky, ``Fast DCT-based image convolution algorithms and application to image resampling and hologram reconstruction,'' Proc. SPIE {\bf 7724}, 77240N (2010).

\bibitem{scale7}
S. Odate, C. Koike, H. Toba, T. Koike, A. Sugaya, K. Sugisaki, K. Otaki, and K. Uchikawa, ``Angular spectrum calculations for arbitrary focal length with a scaled convolution,'' Opt. Express {\bf 19}, 14268--14276 (2011).


\bibitem{scale8}
T. Shimobaba, K. Matsushima, T. Kakue, N. Masuda, and T. Ito,  ``Scaled angular spectrum method,'' Opt. Lett. {\bf 37}, 4128--4130 (2012).

\bibitem{scale9}
T. Shimobaba, T. Kakue, N. Okada, M. Oikawa, Y. Yamaguchi, and T. Ito, ``Aliasing-reduced Fresnel diffraction with scale and shift operations,'' J. Opt. {\bf 15}, 075302(5pp) (2013).

\bibitem{nu}
T. Shimobaba, T. Kakue, M. Oikawa, N. Okada, Y. Endo, R. Hirayama, N. Masuda, T. Ito, ``Non-uniform sampled scalar diffraction calculation using non-uniform Fast Fourier transform,'' Opt. Lett. {\bf 38}, 5130--5133 (2013).

\bibitem{goodman}
J.W.Goodman, ``Introduction to Fourier Optics (3rd ed.),'' Robert \& Company (2005).

\bibitem{cwo}
T. Shimobaba, J. Weng, T. Sakurai, N. Okada, T. Nishitsuji, N. Takada, A. Shiraki, N. Masuda, and T. Ito,  Comput. Phys. Commun. {\bf 183}, 1124--1138 (2012).


\end{thebibliography}







\end{document}